\documentclass{ws-ijmpd}

\newcommand{\beeq}{\begin{equation}}
\newcommand{\eneq}{\end{equation}}
\newcommand{\be}{\begin{eqnarray}}
\newcommand{\ee}{\end{eqnarray}}

\def\d{\delta}

\def\r{\rho}

\def\D{\Delta}

\begin{document}

\markboth{V. K. Onemli} {Probing Cold Dark Matter Cusps by
Gravitational Lensing}

%%%%%%%%%%%%%%%%%%%%% Publisher's Area please ignore %%%%%%%%%%%%%%%
%
\catchline{}{}{}{}{}
%
%%%%%%%%%%%%%%%%%%%%%%%%%%%%%%%%%%%%%%%%%%%%%%%%%%%%%%%%%%%%%%%%%%%%

\title{PROBING COLD DARK MATTER CUSPS\\ BY\\ GRAVITATIONAL LENSING}

\author{V. K. ONEMLI}

\address{Department of Physics and Institute of Plasma Physics, University of Crete,\\
Heraklion, GR-710 03, Greece\\
onemli@physics.uoc.gr}

\maketitle

\begin{history}
\received{Day Month Year}
\revised{Day Month Year}
\comby{Managing Editor}
\end{history}

\begin{abstract}
I elaborate on my prediction that an indirect detection of cold
dark matter (CDM) may be possible by observing the gravitational
lensing effects of the CDM cusp caustics at cosmological
distances. Cusps in the distribution of CDM are plentiful once
density perturbations enter the nonlinear regime of structure
formation. Caustic ring model of galactic halo formation provides
a well defined density profile and a geometry near the cusps of
the caustic rings. I calculate the gravitational lensing effects
of the cusps in this model. As a pointlike background source
passes behind a cusp of a cosmological foreground halo, the
magnification in its image may be detected by present instruments.
Depending on the strength of detected effect and the time scale of
brightness change, it may even be possible to discriminate between
the CDM candidates: axions and weakly interacting massive
particles.
\end{abstract}


\section{Introduction}

\label{sect:intro} Caustics of light\cite{Natural} are generically
the enveloping surfaces of a family of light rays where the
intensity is high. Such concentration of light can burn, hence
these envelopes are called ``caustics.'' The caustic formation in
the propagation of light is a common phenomenon in nature.
Rainbows, scintillation of stars, shimmering of the sea and
dancing patterns of bright lines that appear on the bottom of a
swimming pool are among the most familiar examples. When the
wavefront of sun light enters into water, it gets retarded
proportional to the local thickness of the wavy surface. If the
thickness is small, the amplitude of the wavefront is not affected
much, hence, the density of light rays remains almost uniform just
under the water surface. In a deeper level from the surface,
however, the rays necessarily intersect each other, because the
tiny deviations in the paths of the rays due to the refraction are
amplified. They form cusp-shaped regions bounded by an enveloping
surface of rays. Three light rays hit every point inside this
region, whereas one light ray hits every point outside. The
enveloping surfaces, where the number of light rays jumps by two,
are the fold caustics of light. Cusps occur wherever the folds
merge. The density diverges when a caustic is approached from the
side with extra rays; see Section \ref{sect:light}. Thus, a mild
modulation in the wavefront spontaneously generates a strong
enhancement in intensity some distance away, at caustic locations.
Caustics move about but remain robust under stirrings of the water
surface.

Caustics are common in the propagation of light, because (i)
photons are classically {\it collisionless}, and (ii) flow of
light from a distant (pointlike) source has zero {\it velocity
dispersion}. Caustics of ordinary luminous matter is unusual,
because ordinary matter is normally collisionful. However, the
fall of stars in and out of a galactic gravitational potential
well is collisionless. Many elliptic galaxies are surrounded by
ripples\cite{Malin} in the distribution of light that are
interpreted as caustics of luminous matter.\cite{Ripples} These
ripples provide an {\it existence proof} of discrete flows and
caustics in the infall of ordinary matter which constitutes only
$4\%$ of the content in the Universe. It is believed that $22\%$
of the composition consists of nonluminous, {\it collisionless}
particles with very small {\it primordial velocity
dispersion}.\cite{Spergel} Such particles are called cold dark
matter (CDM). The leading CDM candidates are axions and weakly
interacting massive particles (WIMPs). Galaxies are surrounded by
CDM that keeps falling into the gravitational potential wells of
galaxies from all directions and forms halos around their baryonic
disks. Because the number densities of axions and WIMPs are huge,
their infall onto isolated galaxies is regarded as a continuous
flow. The number of flows, however, is discrete\cite{Ipser,stw1}
at any point in the halo and changes as a function of space and
time. The enveloping surfaces where the number of flows jumps by
two are the CDM caustics. Sharp discontinuities in the CDM density
distribution occur at the locations of the caustics. Existence of
caustics in such a CDM infall\cite{zel}\cdash\cite{tre} can be
taken for granted, once the existence of ripples\cite{Ripples} in
the infall of ordinary mater is given. In a typical galaxy, any
effect one can imagine that may erase CDM caustics, would have
already erased these ripples. Indeed, there are several evidences
for CDM caustics found in the halos of galaxies including the
Milky Way.\cite{Mahdavi}\cdash\cite{IRAS}

\begin{figure}[ht] \centering
\includegraphics[height=5cm,width=11cm]{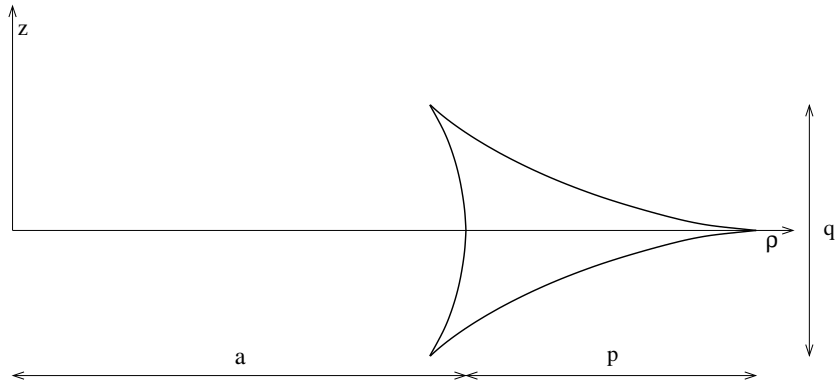}
\caption{Cross-section of a caustic ring in the case of axial and
reflection symmetry. $\rho$ and $z$ are galactocentric cylindrical
coordinates. $p$ and $q$ are the transverse dimensions in the
$\hat{\rho}$ and $\hat{z}$ directions, respectively. $a$ is the
ring radius. $p$ and $q$ are taken of order $a$, for clarity. $p,q
\ll a$ for actual caustic rings.} \label{fig:fig6}
\end{figure}

There are at least two sets of caustics in the halos of isolated
galaxies: outer caustics and caustic
rings.\cite{stw2}\cdash\cite{sing} Outer caustics are simple fold
($A_2$) catastrophes located on topological spheres enveloping the
galaxy. Caustic rings, on the other hand, are closed tubes whose
transverse cross-section is a closed line with three (dual) cusps,
called a ``tricusp.'' One of the cusps, the outer cusp, lies in
the galactic plane, the other two are nonplanar; see
Fig.~\ref{fig:fig6}. The triaxial tricusp is called elliptic
umbilic ($D_{-4}$) catastrophe.

To see the caustic formation\cite{stw2}\cdash\cite{sing} in the
infall of CDM, consider the time evolution of CDM particles that
are about to fall into the potential well of an isolated galaxy at
time $t_1$ for the first time in their history. They form a
topological sphere with radius $R_1(t_1)$ enclosing the galaxy
called the ``turnaround'' sphere at time $t_1$. After falling all
the way through the disk of the galaxy, the particles of this
turnaround sphere form a new sphere which reaches its maximum
radius $R_2(t_2)$ at time $t_2$. $R_2(t_2)<R_1(t_1)$ because of
the deepening of the potential well in the meantime.
 Note also that, due to the growth of the galaxy, the first
turnaround radius $R_1(t_2)$ at time $t_2$ is larger than
$R_1(t_1)$. The sphere continues to oscillate in this way.
Therefore, there are discrete number of flows in and out of the
galaxy at any location in the halo. The local number of flows
increases, as one approaches the galactic center from an arbitrary
direction at a given time. First, in the outermost regions, there
is one flow (particles falling in for the first time), then, at a
certain location, the number jumps to three (particles falling in
for the first time, particles falling out for the first time,
particles falling in for the second time), and then to five, to
seven, and so on, at other certain locations closer to the center.
The boundary ---which is a topological sphere--- between the
region with one (three, five, . . .) flow and the region with
three (five, seven, . . .) flows is the first (second, third, . .
.) outer caustic. Outer caustics are simple fold catastrophes. In
the limit of zero velocity dispersion, the density diverges when
the outer caustics are approached from the side with two extra
flows; see Section \ref{sect:cdmcaus}. To see the formation of
caustic rings with tricusp cross-section, on the other hand, a
closer inspection to the time evolution of the infall sphere is
needed.\footnote{See Ref.~\refcite{sing} for details.} Notice
that, each time the turnaround sphere falls through the galaxy, it
turns itself inside out.\cite{stw2}\cdash\cite{sing} The particles
near the equator of the infall sphere carry too much angular
momentum to reach the central part of the galaxy, and therefore,
just before the sphere turns itself inside out, there exists a
circle of points which are inside the sphere last. This circle is
the locus of outer cusps that lie in the plane of the galaxy (see
Fig.~\ref{fig:fig6}). Hence, caustic rings are located near where
the particles with the most angular momentum in a given inflow
reach their distance of closest approach to the galactic center
before going back out.\cite{sing} Because the CDM infall is
continuous, the caustic ring in space-time associated with the
infall sphere is a persistent feature in space. There is a caustic
ring due to CDM particles falling through the galaxy for the first
time, a caustic ring of smaller radius due to particles falling
through for the second time, and so on. Like outer caustics,
caustic rings are boundary surfaces where the number of flows
jumps by two. Inside the caustic rings there are four flows,
whereas there are two flows outside.

Caustic structure in the dark matter infall may be revealed by
observing their gravitational lensing
effects.\cite{Hogan}\cdash\cite{Vakif} The effects are largest
near the cusps\cite{Vakif} which are common features in the halo
CDM distribution.\footnote{For example, gravity of massive objects
like stars, focus the CDM flow producing a cone-shaped caustic
surface ---whose apex is a cusp--- near the trajectories of
maximum scattering angle.\cite{Wick} This effect can be
generalized considering an isolated galaxy or even an isolated
(super)cluster of galaxies as a massive object, as a whole.} In
this paper, I review the lensing effects near the cusps of caustic
rings.\cite{Vakif} The outline is as follows. In Section
\ref{sect:light}, caustics in the propagation of light are
introduced. They provide a useful analog for understanding the
properties of caustics in the infall of cold dark matter. The
outer caustics and caustic rings of CDM are reviewed in Section
\ref{sect:cdmcaus}. Their gravitational lensing effects are
discussed in Section \ref{sect:lensing}. Conclusions are
summarized in Section \ref{sect:conc}.

\section{Caustics of Light}
\label{sect:light}

When natural processes focus light, the caustics that are
generated have a systematic pattern.\cite{Natural} The
mathematical tool to investigate this systematics is the
catastrophe theory which analyzes the different ways that the
critical points of gradient mappings can merge to make higher
singularities. In optics, the mappings are from an incident wave
front to the space in which the wave is observed; i.e.
three-dimensional (3D) space around the caustic or 2D space of a
screen. The quantity that is extremized is the Fermat's potential
and the critical points correspond to the light rays. The
annihilation of two critical points at a location by coalescence
with each other means loss of two rays contributing to a focus at
that location.

\begin{figure}[ht]
\centering
\includegraphics[height=5cm,width=12cm]{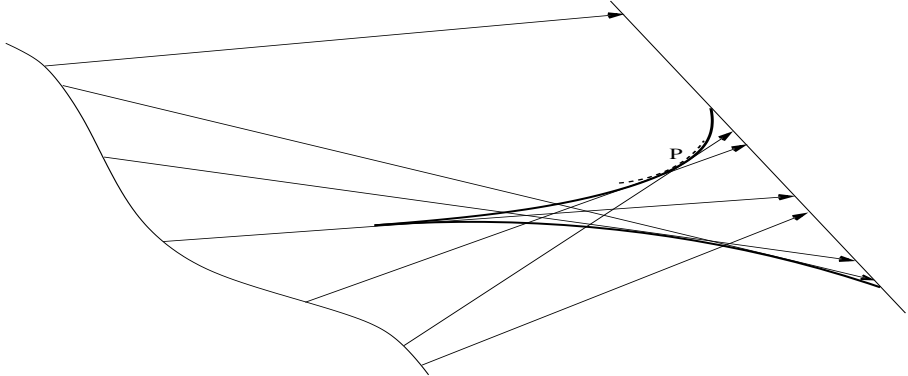}
\caption{Distribution of light rays in the vicinity of a two
dimensional cut of a region bounded by two folds merging at a
cusp. The local curvature radius $R$ of the fold at point $P$
along the tangential direction is that of the circle depicted by
dashed lines.} \label{fig:cusp}
\end{figure}

Caustics spontaneously arise when a forefront object distorts the
propagation of light from a distant (pointlike) source. Although
the light rays are parallel initially, they necessarily cross each
other after the distortion and produce caustics systemized as
folds and cusps. If an observer places his/her eye inside the
region bounded by two folds that merge at a cusp, light rays enter
eye from three distinct directions, each corresponding to an image
of the source. Every point outside the region, on the other hand,
is hit by one ray only. A 2D cross-section of such a region is
depicted in Fig.~\ref{fig:cusp}.
Thus, as the point of observation crosses a fold caustic surface,
the number of images jumps by two. If the eye moves from the
inside across one of the folds, two of the images approach each
other increasing in brightness, coalesce at the fold and then
disappear outside. Meanwhile, the third image remains pretty much
the same. If the eye moves towards the other fold, the third image
coalesces with one of the other two images at the fold and then
disappear. The closer the eye is to the cusp, the closer the two
images are. The three images merge at the cusp. Because rays are
more concentrated near the cusps, the intensity is largest there;
see Fig.~\ref{fig:cusp}. When a fold is approached from the side
with two extra rays at every point, the density rises as
$\sigma^{-1/2}$ in the vicinity of the fold, where $\sigma$ is the
distance to the fold in the direction of approach. This can be
seen as follows. The rays tangent to a small segment of a smooth
caustic fold at a point $P$ are uniformly
distributed\cite{Natural} along it; see Fig.~\ref{fig:cusp}. The
corresponding length of the segment $\simeq 2\sqrt{2R\sigma}$
where $R$ is the local curvature radius of the fold at $P$ along
the tangential direction. Therefore, in the caustic neighborhood,
the number of rays $N$ passing between the observer and the fold
is proportional to $2\sqrt{2R\sigma}$. Hence, the number density
$dN/d\sigma\sim\sigma^{-1/2}$. Because caustics of light form
spontaneously in rather general circumstances and retain their
identity under perturbations, they are generic and structurally
stable.

\section{Caustics of Cold Dark Matter}
\label{sect:cdmcaus}

CDM infall onto isolated galaxies {\it intrinsically} has the
necessary conditions for the caustic formation noted in
Section~\ref{sect:intro}: (i) the particles interact only weakly,
and (ii) the primordial velocity dispersions $\delta v$ of the CDM
candidates are of a very small order. For the axion infall, the
primordial velocity dispersion at time $t$, is of
order\cite{stw2,Chang} $\delta v_a (t) \sim 10^{-12} \,{\rm
km/s}\, \left( {10^{-5} {\rm eV}/ m_a}\right)^{5/6}\left({t_0/
t}\right)^{2/3}$, where $t_0$ and $m_a$ are respectively the
present age of the Universe and the axion mass. For the WIMP
infall, $\delta v_\chi(t)\sim 10^{-6}\,{\rm km/s}\,\left({{\rm
GeV}/m_\chi}\right)^{1/2}\left({{\rm
MeV}}/{T_D}\right)^{1/2}\left({t_0}/{t}\right)^{2/3}$, where $T_D$
and $m_\chi$ are respectively the decoupling temperature and the
mass of the WIMPs.\cite{stw2,sing} (Detailed derivations of
$\delta v_a$, and $\delta v_\chi$ ---including the possible effect
of reheating by the $e^-e^+$ annihilations--- can also be found in
Ref.~\refcite{Thesis}). Thus, stable and generic caustic formation
in the CDM infall is inevitable, once density perturbations enter
the nonlinear regime. In this section, I review the properties of
the minimal caustic structure\cite{stw2}\cdash\cite{sing} that
must occur in galactic halos: outer caustics and caustic rings.

\subsection{Outer caustics}

\label{sub:outer} Outer caustics are topological spheres
enveloping the galaxies. In catastrophe theory, they are known as
simple fold ($A_2$) catastrophes. \begin{figure}[ht] \centering
\includegraphics[height=6cm,width=9.5cm]{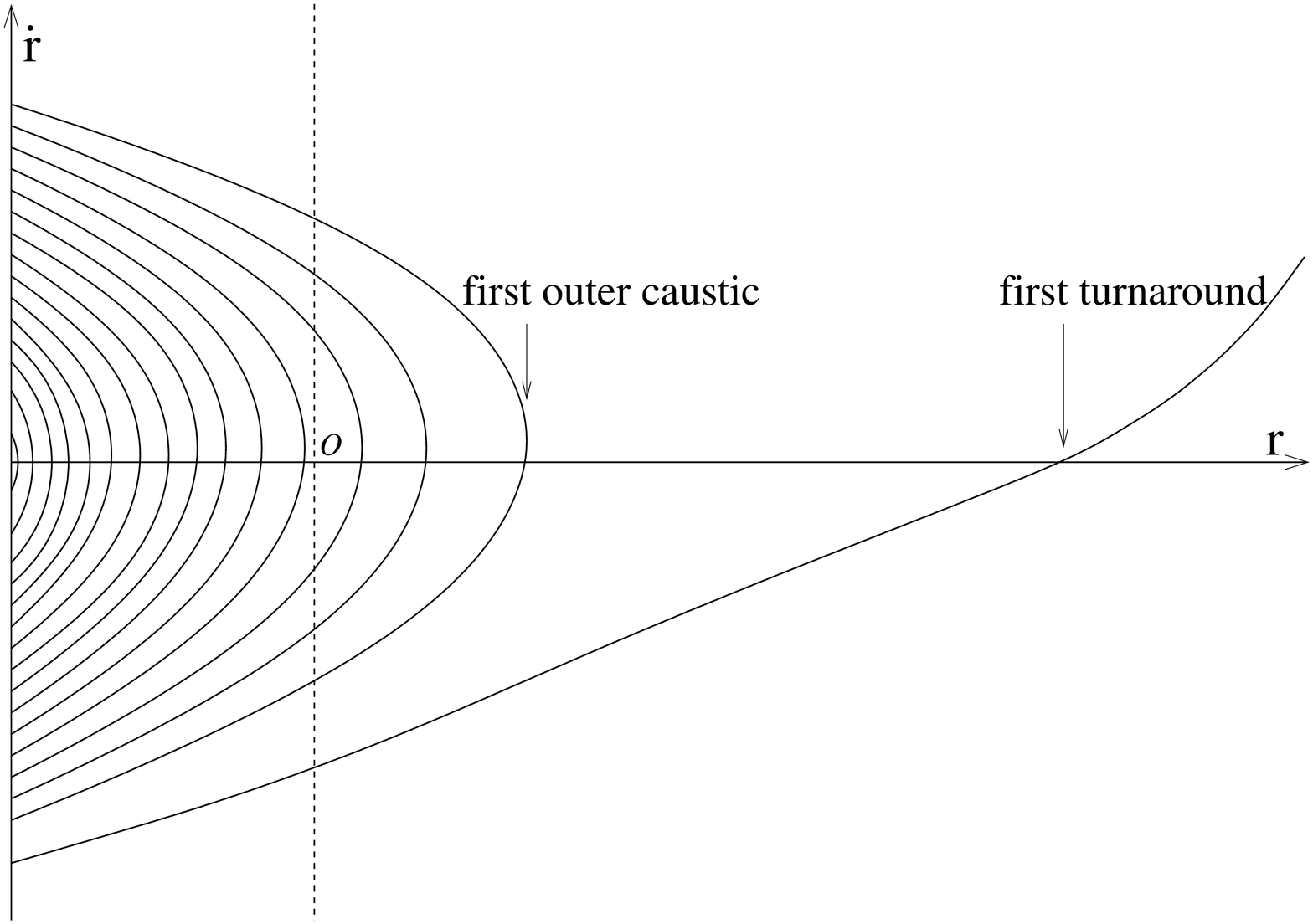}
\caption{Snapshot of the phase space of CDM particles in a
galactic halo. $r$ is galactocentric distance and $\dot r$ is
radial velocity. The solid line indicates the location of the
particles. The dotted line corresponds to observer position. Each
intersection of the solid and dotted lines corresponds to a CDM
flow at the observer's location $O$. Caustics occur wherever the
phase space line folds back.} \label{fig:phase}
\end{figure}To derive the properties of the
outer caustics, it is instructive to discuss their formation in
the phase space point of view.\cite{stw2}\cdash\cite{sing} The
small velocity dispersion $\delta v$ means that CDM particles lie
on a very thin 3D sheet in 6D phase space. The thickness of the
sheet is the $\d v$ of the particles. The flows can be described
in terms of the evolution of this sheet.\cite{Ipser,stw2} Figure
\ref{fig:phase} depicts two dimensional cut $(r, \dot r)$ of the
phase space, $r$ being the radial coordinate and $\dot r$ being
the radial velocity.\cite{stw2} Before density perturbations enter
the nonlinear regime of structure formation, there is only one
flow (i.e. a single value for velocity) at a typical location in
physical space. When a large overdensity enters the nonlinear
regime, the particles in the vicinity of the overdensity fall back
onto it. This implies that the phase space sheet winds up in phase
space wherever overdensity grows. The outcome of this process is
an odd number of discrete flows at any point in a galactic halo;
see Fig.~\ref{fig:phase}. Between the first
turnaround\footnote{``Turnaround'' refers to the moments in a
particle's history when it has zero radial velocity with respect
to the galactic center.} radius and the location where the phase
space sheet folds back for the first time
---near\footnote{The $n$-th fold is located near the $n\!+\!1$-th
turnaround radius; see Fig.~\ref{fig:phase}. If the flow were
stationary the $n$-th fold would be located exactly at the
$n\!+\!1$-th turnaround radius.} the second turnaround radius---
there is one flow (associated with particles falling in through
the galaxy for the first time). Between the first and second folds
there are three flows (associated with particles falling in
through the galaxy for the first and second times, and with
particles falling out of the galaxy for the first time). Between
the second and third folds there are five flows and so on.
(Wherever the phase space sheet folds back, the number of flows in
physical space jumps by two naturally). The surface in physical
space which envelopes the region with the two extra flows,
associated with the corresponding fold of the phase space sheet,
is a CDM caustic. At the fold, the phase space sheet is tangent to
velocity space, particles pile up there and the physical space
density diverges as the inverse square root of the distance
$\sigma$ to the caustic, when it is approached from the side with
the two extra flows. This can be shown as follows.\cite{Thesis}
The 2D phase space mass density near a fold may be represented as
${d^2M}/{d\sigma d\dot\sigma}=
\textsf{M}{\Theta}(\sigma)\delta(\sigma-(C\dot{\sigma})^2)$, where
$C$ and $\textsf{M}$ are positive numbers; see
Fig.~\ref{fig:phase}. The direction of $\sigma$ coordinate is
chosen pointing inward. The density is the integral of the phase
space density over velocity space \be d(\sigma )= \textsf{M} \int
d\dot\sigma\frac{\Theta(\sigma)}{2C}[\delta(\dot \sigma
-{{\sigma}^{1\over2}/{C}})+ \delta(\dot \sigma
+{{\sigma}^{1\over2}/{C}})]\sigma^{-{1\over2}}
=A\Theta(\sigma)\sigma^{-{1\over2}}\; ,\ee where $A$ is called the
fold coefficient. To estimate\cite{lensing} $A$, consider the time
evolution of CDM particles which are falling out of a galactic
halo for the $n$th time and then falling back in. Let $R_n$ be the
$n$th turn-around radius. We assume that the rotation curve of the
galaxy is flat near $r=R_n$ with time-independent rotation
velocity $v_{\rm rot}$. Then, using the force per unit mass
$\vec{f}(r)=-{v^2_{\rm{rot}}}\hat r/r$ and conservation of energy
per unit mass $E=\frac{1}{2}\dot{\vec{r}}\,^2+V(r)$, the
gravitational potential $V(r)$ is calculated as $
V(r)=v^2_{\rm{rot}}\ln\left(r/R_n\right)+E$. Thus, the particles
have a trajectory $\vec{r}(t)$ such that
$|{\dot{\vec{r}}}|=v_{\rm{rot}}\sqrt{2
\ln\left({R_n}/{r}\right)}$. Here, we neglected the angular
momentum of the particles, since at the turaround the particles
are far from their distance of closest approach to the galactic
center. Finally, we assume that the flow is stationary. This means
the number of particles flowing per unit solid angle and per unit
time, ${dN}/{d\Omega dt}$, is independent of $t$ and $r$, and
hence, the caustic is located exactly at the $(n+1)$th turnaround
radius $R_n$. The mass density of particles, $d_n(r)$, follows
from the equality: $2dM=d_n(r) r^2 d\Omega\,\dot{r}dt$. The factor
of two appears because there are two distinct flows, out and in.
Near the $n$th outer caustic where $r=R_n-\sigma$,
$\ln{\left(R_n/r\right)}\simeq{\sigma}/{R_n}$. Therefore,
$d_n(r)\simeq A_n\Theta(\sigma)\sigma^{-1/2}$ with
$A_n=\sqrt{2}({dM}/{d\Omega dt}|_n)v^{-1}_{\rm{rot}}R^{-3/2}_n$.
In a self-similar infall, $R_n$ and $A_n$ are
estimated\cite{lensing} as\be \!\!\!\!\!\!\!\!\{R_n: n=1,2,.\; .\;
.\} \!&\simeq&\! (240,~120,~90,~70,~60,.\; .\; .)~{\rm kpc}
\cdot\!\left(\frac{v_{\rm{ rot}}}{220\,{{\rm
km/s}}}\right)\! \left(\frac{0.7}{h}\right) \; , \label{Rn}\\
\!\!\!\!\!\!\!\!\{A_n :n=1,2,.\; .\; .\} \!&\sim&\!
(2,~2,~2,~3,~3,.\; .\; .)\cdot {\rm \frac{10^{-5}
\,gr}{cm^2\,kpc^{1/2}}} \cdot\!\left(\frac{v_{\rm{
rot}}}{220\,{{\rm km/s}}}\right)^{1\over2}\!
\left(\frac{h}{0.7}\right)^{3\over2}\; . \label{An}\ee

Evidence for the first outer CDM caustic was recently
found\cite{Mahdavi} in the NGC 5846 group of galaxies. The plot of
the surface number density of galaxies was inferred to mark the
first outer caustic of radius $R_1$ by an abrupt density drop and
a transition from a large $\delta v$ to a very small $\delta v$,
as predicted by the CDM infall models.\cite{FG}

\subsection{Caustic rings}

\label{sub:rings} Caustic rings were first noted as topological
circles in Ref.~\refcite{stw2}, but their properties were left
unexplored there. In Ref.~\refcite{sing} they were studied more
precisely as closed tubes whose transverse cross-section is a
$D_{-4}$ catastrophe; see Fig.~\ref{fig:fig6}. Treatment of
caustic rings, from the catastrophe theory point of view, is given
in Ref. \refcite{Vakif}. Caustic rings are made up of three fold
caustics joining at three topological circle of dual cusp
($A_{-3}$) catastrophes. Inside (outside) the caustic rings, there
are four (two) flows. For simplicity, we consider caustic rings
which are axially symmetric about the $z$ direction as well as
reflection symmetric with respect to $z=0$ plane. The CDM flow in
that case is effectively two dimensional. In galactocentric
cylindrical coordinates, the flow near such a ring is described
by:
\be\rho(\chi_1, \chi_2)&=&a+{\chi_1}^2-{\chi_2}^2 \; ,\nonumber\\
z(\chi_1, \chi_2)&=&2\zeta\,(\sqrt{p}-{\chi_1}){\chi_2} \;
,\label{flownewz} \ee where
${\chi_1}=\sqrt{\frac{u}{2}}\,(\tau_0-\tau),\;\;\;
{\chi_2}=\sqrt{\frac{s}{2}}\,\alpha$. $\rho$ is distance to the
$z$-axis. The parameters $\tau$ and $\alpha$ label the particles
in the flow. $\tau$ is the time a particle crosses the $z=0$
plane. $\alpha$ is the declination of the particle, relative to
the $z=0$ plane, when it was at last turnaround. $a$ is the radius
of the caustic ring. The constants $b, s, u$ and $\tau_0$ are the
other characteristics. $\zeta\equiv\frac{b}{\sqrt{us}}$, and
$p\equiv\frac{1}{2}u\tau_0^2$ is the longitudinal dimension of the
caustic ring cross-section.\footnote{If $\zeta=1$, the caustic
ring cross-section is a triaxial tricusp.} The number density of
particles in physical space
\begin{equation}
d(\rho,z,t) = {1\over 2\pi\rho}~\sum_{j=1}^{n(\rho,\; z,\;
t)}~{d^2N\over d{\chi_1} d{\chi_2}}~(\vec{\chi}_j(\rho,\; z,\;
t))~{1\over\mid D_2(\vec{\chi})\mid}\Biggl|_{\vec{\chi} =
{\vec{\chi}}_j(\rho,\; z,\; t)}\; , \label{dens}
\end{equation}
where $\vec{\chi}\equiv({\chi_1}, {\chi_2})$, and
$\vec{\chi}_j\equiv({\chi_1}, {\chi_2})_j$, with $j = 1.\; .\;
.n$, are the solutions of $\rho = \rho ({\chi_1}, {\chi_2}; t)$
and $z= z({\chi_1}, {\chi_2}; t)$. The number of distinct
solutions (flows) $n$, is a function of the location in space
$\rho$, $z$, and time $t$. For the flow near caustic rings, the
determinant of the Jacobian matrix \beeq D_2=
-4\zeta\left[\left({\chi_1}-\sqrt{p}/2\right)^2
+{\chi_2}^2-{p}/{4}\right]\; .\eneq Caustic rings occur where
$D_2$ vanishes, i.e. where the map $(\chi_1,
\chi_2)\rightarrow(\rho, z)$ is singular. Hence, at the caustic,
the density diverges in the limit of zero velocity dispersion. In
the parameter space of $\chi_1$ and $\chi_2$, the curve for which
$D_2=0$ is a circle centered at $({\chi_1} ,
{\chi_2})=({\sqrt{p}}/{2} , 0)$ with radius ${\sqrt{p}}/{2}$. A
one-parameter representation of this critical circle can be
given\cite{Vakif} as: \be
{\chi_1}=\frac{\sqrt{p}}{2}\left(1+\cos\psi\right),
\;\;{\chi_2}=\frac{\sqrt{p}}{2}\sin\psi\; ,\label{causticpar}\ee
where the angular variable $\psi\in [0, 2\pi]$. Substituting Eq.~
(\ref{causticpar}) into Eqs.~(\ref{flownewz}) yields the equations
that describe the cross-section of the caustic ring in the $(\rho
, z)$-plane \beeq
\rho(\psi)=a+\frac{p}{2}\cos\psi(1+\cos\psi),\;\;
z(\psi)=\zeta\frac{p}{2}\sin\psi(1-\cos\psi)\; . \label{zpsi}
\eneq Figure \ref{fig:fig6} shows a plot of this cross-section.
The cusps occur at $\psi=0,2\pi/3,4\pi/3$. In a self-similar
infall, the caustic ring radii are estimated\cite{cr} as\beeq
\{a_n: n=1, 2,.\; .\; .\}\simeq(39,~19.5,~13,~10,~8,.\; .\; .){\rm
kpc}\cdot\!
\left(\frac{j_{\rm{max}}}{0.27}\right)\!\left(\frac{0.7}{h}\right)\!
\left(\frac{v_{\rm{rot}}}{220\,{\rm km/s}}\right)\; , \label{a_n}
\eneq where $j_{\rm {max}}$ is a parameter proportional to the
amount of angular momentum that the particles have, $h$ is the
present Hubble constant in units of 100 km/(s Mpc) and $v_{\rm
rot}$ is the rotation velocity of the galaxy. Because the caustic
rings lie close to the galactic plane, they cause bumps in the
rotation curve, at the locations of the rings. In a study of 32
extended and well-measured external galactic rotation curves,
evidence was found for the law given in Eq.~(\ref{a_n}).\cite{KS}
In the rotation curve of the Milky Way, the locations of eight
sharp rises fit the prediction of the self-similar model at the
3$\%$ level.\cite{IRAS} Moreover, a triangular feature in the
Infrared Astronomy Satellite (IRAS) map of the Milky Way plane was
interpreted as the imprint on baryonic matter of the fifth caustic
ring.\cite{IRAS} This nearby ring is 55 pc away from us. The
feature is correctly oriented with respect to the galactic plane
and the galactic center. Its location is consistent with the bump
between 8.28 kpc and 8.43 kpc in the rotation curve. The
probability that the coincidence in position of the feature with a
rise in the rotation curve is less than $10^{-3}$.

The density profile near the caustic ring along the direction
$\hat\sigma$ perpendicular to the surface is $d(\psi, \sigma) =
A(\psi)\Theta(\sigma)\sigma^{-1/2}$, where the fold coefficient
\beeq A(\psi)= \frac{d^2M}{d\Omega dt}\sqrt{\frac{2\zeta}{p}}
\frac{\cos{\alpha(\psi)}}{bC(\psi)\rho(\psi)}\; ,
\label{Apsi}\eneq with\cite{Vakif}\beeq
C(\psi)=\sqrt{\left|(1+2\cos\psi)\tan\frac{\psi}{2}\right|
\sqrt{(1+\cos\psi)^2+\zeta^2\sin^2\psi} }\; . \eneq The fold
coefficients are estimated as a function of location on the ring
as
 \be\{A_n(\psi)\!:\! n\!=\!1, 2,.\; .\;
.\}\!\sim\! (5, 6, 6, 7, 8,.\;.\;.)\cdot
10^{-4}\cdot\frac{{\mathcal F}_n(\psi)\,{\rm gr}}{\rm
cm^2\,kpc^{\frac{1}{2}}}\nonumber\\
\times\left(\frac{0.27}{j_{\rm{max}}}\right)^{\frac{3}{2}}
\!\!\left(\frac{h}{0.7}\right)^{\frac{3}{2}}\!\!
\left(\frac{v_{\rm{rot}}}{220\, {\rm
{km}/{s}}}\right)^{\frac{1}{2}}\; , \label{Anring} \ee
where\cite{Vakif} \beeq {\mathcal F}_n(\psi)=
\frac{[1+\frac{p_n}{2a_n}\cos\psi(1+\cos\psi)]^{-1}}
{\sqrt{|(1+2\cos\psi)\tan\frac{\psi}{2}|
[(1+\cos\psi)^2+\zeta^2_n\sin^2\psi]^{\frac{1}{2}}} }\;\;
.\label{Fn}\eneq Due to the symmetry of the tricusp with respect
to the $z=0$ axis, we have ${\mathcal F}_n(\vartheta)={\mathcal
F}_n(-\vartheta)$ and ${\mathcal
F}_n(\frac{2\pi}{3}\pm\vartheta)={\mathcal
F}_n(\frac{4\pi}{3}\mp\vartheta)$. For the typical values ${p_n}=
0.1~ a_n$ and $\zeta_n= 1$, near the outer cusp ${\mathcal
F}(\vartheta)\simeq 0.525~ \vartheta^{-\frac{1}{2}}$. Near the
nonplanar cusps ${\mathcal F}(\frac{2\pi}{3}\pm\vartheta)\simeq
0.585~ \vartheta^{-\frac{1}{2}}$. This $\vartheta^{-\frac{1}{2}}$
divergence of the fold coefficients at the cusps is cut off when
$\delta v\neq 0$, because, the location of the caustic surface
gets smeared over some distance $\delta x$ and the cusps are
smoothed out. For CDM caustics in galactic halos, $\delta x$ and
$\delta v$ are related\cite{cr,Hogan,lensing} as $\delta x \sim
{R~\delta v/ v}$ where $v$ is the order of magnitude of the
velocity of the particles in the flow and $R$ is the distance
scale over which that flow turns around. For a galaxy like our
own, $v = 500$ km/s and $R = 200$ kpc are typical orders of
magnitude. As a result of primordial velocity dispersion, axion
and WIMP caustics in galactic halos are typically smeared over
$\delta x_a \sim 6\cdot 10^{4}\, {\rm km} \left({10^{-5} {\rm eV}
/ m_a} \right)^{5/6}$ and $\delta x_\chi
 \sim 10^{10}\, {\rm km} \left({{\rm GeV}
/ m_\chi}\right)^{1/2}$, respectively.  However, a CDM flow may
have an effective velocity dispersion $\d x_{\rm eff}$ which is
larger than its primordial velocity dispersion. Effective velocity
dispersion occurs when the sheet on which the dark matter
particles lie in phase space is wrapped up on sub-scales that are
small compared to the galaxy as a whole. Little is known about the
size of the $\d x_{\rm eff}$ in galactic halos. However, the
widths of the bumps at the ring locations in the rotation curve of
the Milky Way and the sharpness of the edges of the triangular
feature in the IRAS map were used to deduce an upper bound of
15-20 pc on $\d x_{\rm eff}$ {\it at most}.\cite{IRAS}

I use the primordial smearing distances $\delta x_a$ and $\delta
x_\chi$ to estimate the upper limits for the fold coefficient and
for the lensing effects at the cusps of caustic rings as follows.
I set the $\delta v$ of the particle species equal to zero, and
then calculate the fold coefficients and the lensing effects
(Sect. \ref{sect:lensing}) on the caustic surface at a point whose
transverse distances $\D\r$ and $\D z$ to the nearby cusp
---that would occur in the limit $\delta v=0$--- are slightly
greater than the primordial smearing out distance $\delta x_a$ or
$\delta x_\chi$. These points correspond to the locations that are
respectively $\Delta\psi=\frac{\pi}{7500}$ and
$\Delta\psi=\frac{\pi}{75}$ radian away from the cusps in the
axionic and WIMPic cases, respectively. Similarly, I use the upper
bound for the smearing distance $\d x_{\rm eff}$ to estimate the
lower limits: I choose a point which is about the maximum smearing
distance $\d x_{\rm eff}$ away from the nearby cusp and calculate
the fold coefficients and the lensing effects there. Such points
differ by $\D\psi=\frac{\pi}{7.5}$ radians from the cusp
locations. The calculations\cite{Vakif} show that, for the axion
caustic rings, the factor ${\mathcal F}(\psi)$ may change the
estimates given for the $A_n$ in Eq.~(\ref{Anring}) between 0.8
and 26 times at the outer cusp, and between 0.9 and 29 times at
the nonplanar cusps. The upper bounds of $A_n$ for the WIMP
caustic rings, on the other hand, are ten times less then the
upper bounds for the axionic rings.

\section{Gravitational Lensing by the CDM Caustics}
\label{sect:lensing}

Matter curves the spacetime. As a result, light rays coming from a
background source are deflected due to the presence of a forefront
object. This effect is known as gravitational lensing. Choosing
the $y$-axis in the direction of propagation of light, the angular
shift in the position of the source $\vec{\xi}_{S}\equiv(\xi_{Sx},
\xi_{Sz})$ due to a lens is given by $\D\xi\equiv\vec{\xi}_{
I}-\vec{\xi}_{S}=\vec{\nabla}_{\xi_{ I}}\Phi(\vec{\xi}_{ I})$,
where ${\vec{\xi}}_{I}\equiv(\xi_{Ix}, \xi_{Iz})$ is the position
of the image with lens present. The 2D potential $\Phi$ is solved
from the Poisson equation $\nabla^2_{\xi_{ I}}\Phi= 2
{{\Sigma}/{\Sigma_c}}$,  where column density $\Sigma$ and the
critical surface density $\Sigma_c$ are defined as:
\be\Sigma(\vec{\xi}_{ I})=\int dy\, d(D_L\xi_{ Ix},y,D_L\xi_{
Iz})\; ,\nonumber\ee \beeq\Sigma_c={{c^2D_{S}} \over{4\pi G D_L
D_{LS}}}=0.347 \, {\rm{\frac{g}{cm^2}}} \left({{D_{S}}\over{D_L
D_{LS}}}\,{\rm{Gpc}} \right)\, . \label{scri} \eneq Here, $D_L$
and $D_S$ are the distances of the observer to the lens and to the
source, respectively. $D_{LS}$ is the distance of the source to
the lens. The image distortion and magnification are given by the
Jacobian matrix of the map $\vec\xi_S(\vec\xi_I)$. Because
gravitational lensing does not change surface brightness, the
magnification ${\mathcal{M}}$ is the ratio of image area to source
area. Hence,
${\mathcal{M}}=1/{|\det{(\partial\xi_{Si}/\partial\xi_{Ij})}|}$,
which yields, to first order, ${\mathcal{M}}=
1+2{\Sigma/\Sigma_c}$. Gravitational lensing has proven useful in
revealing the massive compact halo objects (MACHOs) in galaxies
and constraining the mass distribution in galaxy clusters. For
some recent interesting lensing applications see
Refs.~\refcite{KK}. Dark matter caustics have calculable lensing
signatures.\cite{Hogan}\cdash\cite{Vakif} The caustic ring
model\cite{stw2}\cdash\cite{sing} precisely predicts the density
and the geometry of the CDM distribution in caustic neighborhood.
In Ref.~\refcite{lensing}, we derived the lensing equations for
the line of sights that are parallel to the galactic plane of a
caustic ring and near tangent to its surface in some specific
cases. In Ref.~\refcite{Vakif}, I obtained the lensing equations
for the same line of sights, but near tangent to a caustic ring
{\it anywhere} on the surface and estimated the lensing effects
near the cusps of axion and WIMP caustic rings. Here, I summarize
these effects.

\begin{figure}[ht] \centering
\includegraphics[height=10.5cm,width=7.5cm]{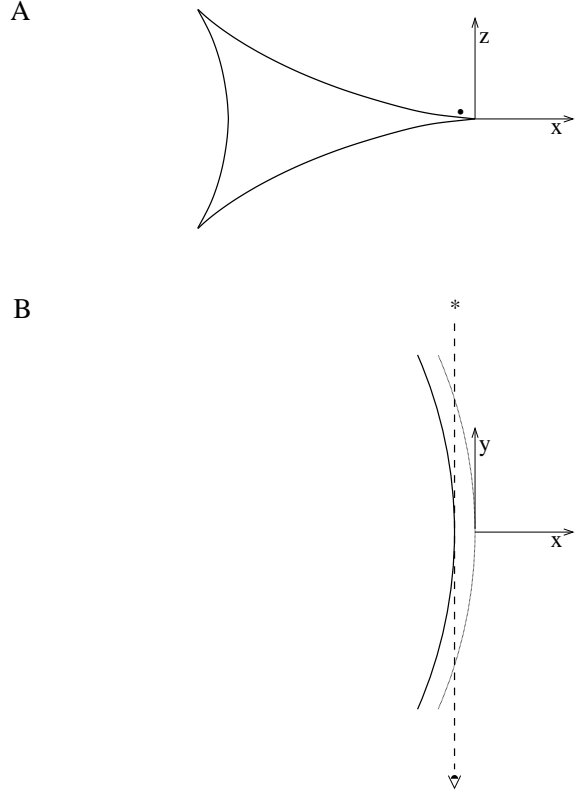}
\caption{Lensing by a caustic ring for a line of sight near
tangent to the surface at a given point $\psi_*$ close to the
outer cusp. The line of sight lies in the $z=z(\psi_*)$-plane. A)
Side view in the direction of the line of sight. The latter is
represented by the dot near $x=z=0$. B) Top view.}
\label{fig:lensingbycusp}
\end{figure}

The gravitational lensing effects of a caustic are largest when
the line of sight is near tangent to the surface, because the
contrast in column density $\Sigma$ is largest there. For such
line of sights, the associated fold of outer caustics, in general,
and of caustic rings, in the regions where $0<\psi<2\pi/3$ and
$4\pi/3<\psi<2\pi$, are curved towards the side with the two extra
flows; see Fig.~\ref{fig:lensingbycusp}.\footnote{If the line of
sight is near tangent to the ring surface at
$\psi\in[\frac{2\pi}{3}, \frac{4\pi}{3}]$, the associated fold is
curved away from the side with the two extra flows. The expected
effects are not as large in that
case.\cite{lensing,Vakif}} In these cases, the image shift and magnification are
given\cite{lensing} as \be \Delta\xi=\xi_{ I}-\xi_{ S} =
\eta\,\xi_{ I}\,\Theta{(-\xi_{ I})}\; ,\;\;{\mathcal{M}}={{d\xi_{
I}}\over{d\xi_{S}}} = 1 + \eta\;\Theta{(-\xi_I)} + O(\eta^2)\; ,
\label{Mm} \ee where \beeq \eta = {{2\pi
A\sqrt{2|R|}}\over{\Sigma_c}} \; .\label{ep} \eneq Here $R$ is the
curvature radius of the surface along the direction associated
with the chosen line of sight. Hence, when the line of sight of a
moving source crosses the caustic, the component of its apparent
velocity perpendicular to the surface and the magnification of the
image changes abruptly. Both affects are of order $\eta$.

The $\eta_n$ for the outer caustics are estimated,\cite{lensing}
using Eqs.~(\ref{Rn}) and (\ref{An}) in Eq.~(\ref{ep}), as \be
\{\eta_n\, : n=1, 2,.\; .\; .\}\sim (7,~6,~6,~5,~5,.\; .\; .)\cdot
10^{-3}\nonumber\\ \times \left(\frac{D_L D_{LS}}{D_S~{\rm
Gpc}}\right) \left(\frac{v_{\rm {rot}}}{220 {\rm km/s}}\right)
\left(\frac{h}{0.7}\right)~~~\ . \label{epe} \ee Lensing effects
are always maximum when the lens is situated half-way between the
source and the observer. Also, $D_S$ should be as large as
possible to get the largest effects. Therefore, in my estimates, I
will assume that the source is at cosmological distances, e.g.
$2D_L=2D_{LS}=D_S=1$ Gpc. For the outer caustics, this yields a
magnification of order $10^{-3}$ which can not be detected by
present instruments. However, the images of extended sources may
be modified in recognizable ways.\cite{Hogan,lensing} In
particular, if a straight jet makes angle $\alpha$ with the
normal, it appears bent by an angle $\eta\sin(2\alpha)/2$  where
its line of sight crosses a fold of an outer
caustic.\cite{lensing}

For the caustic rings, $A$ is given in Eq.~(\ref{Apsi}). Along the
chosen line of sights, the curvature radius of the associated
fold\cite{Vakif}\be R=-\zeta^{-1}{\rm Sign}(\cos\psi-\cos
2\psi)[a+\frac{p}{2}\cos\psi(1+\cos\psi)](\zeta^2
+\cot^2\frac{\psi}{2})^{\frac{1}{2}}\; .\label{rf}\ee
Therefore\cite{Vakif} \beeq
\eta=\frac{4\pi}{\Sigma_c}\frac{d^2M}{d\Omega
d\tau}\frac{\cos{\alpha(\psi)}}{b\sqrt{ap}} {\mathcal G}(\psi)\;
,\label{ETA}\eneq where \beeq {\mathcal G}(\psi)=
\sqrt{\frac{|\csc\psi|}
{[1+\frac{p}{2a}\cos\psi(1+\cos\psi)]|(1+2\cos\psi)\tan\frac{\psi}{2}|}}\;\;
. \eneq The $\eta_n(\psi)$ for the rings are estimated\cite{Vakif}
as \be \{\eta_n(\psi): n=1,2,. . .\}\sim (7, 6, 6, 5, 5,. .
.)\cdot 10^{-2} {\mathcal G}_n(\psi)\frac{D_L\,D_{LS}}{D_S\,{\rm
Gpc}}\nonumber\\
\times\left(\frac{0.27}{j_{\rm{max}}}\right)\!\!
\left(\frac{h}{0.7}\right)\!\! \left(\frac{v_{{\rm
rot}}}{220\,{\rm km/s}}\right)\; .\label{etan}\ee In the limit of
zero velocity dispersion ${\mathcal G}(\psi)$ diverges at the
cusps. ${\mathcal G}(\vartheta)\simeq 0.8~\vartheta^{-1}$, whereas
${\mathcal G}(2\pi/3-\vartheta)\simeq 0.6~\vartheta^{-1/2}$.
Hence, the effects are large near the cusps, in particular near
the outer cusp. For the line of sights near tangent to a
cosmological ring surface in the region $-\pi/27<\psi<\pi/27$, the
magnification is more than $10\%$. Effects of that order can be
observed by present instruments.\cite{Popowski}

The lensing effects of the cusps of axion and WIMP caustic rings
may be estimated\cite{Vakif} by turning the $\delta v$ of the
particle species on, as described in Sect. \ref{sub:rings}. At the
outer cusp of a cosmological axion caustic ring, the range of
$\eta$ is constrained between $\eta(\pm\frac{\pi}{7.5})\simeq
2.9\cdot 10^{-2}$ and $\eta(\pm\frac{\pi}{7500})\simeq 27.9$. This
implies a magnification between $3\%$ and $2800\%$. At the outer
cusp of a cosmological WIMP caustic ring, on the other hand, the
range is constrained between $\eta(\pm\frac{\pi}{7.5})\simeq
0.029$ and $\eta(\pm\frac{\pi}{75})\simeq 0.28$. This implies a
magnification between $3\%$ and $28\%$ at the outer cusp. The
lower bound of $\eta$ at the nonplanar cusps of cosmological
caustic rings is $\eta(\frac{2\pi}{3}-\frac{\pi}{7.5})\simeq
0.016$. For a cosmological axion caustic ring, the upper bound is
$\eta(\frac{2\pi}{3}-\frac{\pi}{7500})\simeq 0.46$. Thus, the
magnification can be between $2\%$ and $46\%$ at the nonplanar
cusps. (Twenty five MACHOs were discovered\cite{Popowski} by
magnifications that range between $12\%$ and $46\%$). For a
cosmological WIMP caustic ring, on the other hand,
$\eta(\frac{2\pi}{3}-\frac{\pi}{75})\simeq 0.046$, hence the
magnification can be between $2\%$ and $5\%$ at the nonplanar
cusps.

Pointlike background sources may probe the CDM caustics, as they
cross behind the cusps of a foreground halo. The time scale
$\delta t$ for the brightness to change is about the transit time
of a background source across the thickness (i.e. smearing out
distance) $\delta x$ of the caustic edges. For axion caustics
$\delta t_a$ is about an hour, whereas for WIMP caustics $\delta
t_\chi$ is about a year.\cite{Hogan} A strategy to detect the
caustics may be a MACHO style experiment: monitoring large number
of pointlike background sources behind the disks of foreground
halos and watching for their edge crossings near the cusps. The
number of outer caustics across the face of a halo is about the
number of orbits (discrete flows) since it is
formed,\cite{Ipser,Hogan} say $N\approx 10$ to $100$. The number
of caustic rings are of the same order, because, the very same
flows that generate the outer caustics, generate the caustic
rings.\footnote{The sharp rises in the rotation curve of the Milky
Way imply the existence of 13 caustic rings.\cite{IRAS}} Caustic
ring radii $a_n\sim 40/n$ kpc. Each background source traverses
such a distance scale about $10^8/n$ years. So monitoring $10^8$
stars for a year will show about 20 to 200 crossings of the ring
edges. Only $6\%$ of the ring surface (near the cusps) can magnify
the sources more than $10\%$, hence about 1 to 10 cusp crossings
may be detected in a year. Note that I estimated the prospects
considering the minimal cusp structure predicted by the caustic
ring model. Cusps are generic and common throughout the halos.
More background sources may easily pass behind the generic cusps
of a foreground halo.

\section{Conclusions}
\label{sect:conc}

Caustics are common features in the propagation of light, because
photons are collisionless and flow of light from a point source
has zero velocity dispersion. In the occasions where these two
conditions are satisfied, caustics in the flow of the ordinary
luminous matter are also observed. Cold dark matter (CDM) flow
onto isolated galaxies has both of the properties; the particles
interact only weakly and the flow has a tiny velocity dispersion.
Therefore, caustic formation in the halos of CDM is inevitable.
Gravitational lensing effects of the cusps at cosmological
distances may be detected by present instruments. Caustic ring
model of galactic halos predicts the geometry and density near the
cusps of the rings. I derived the lensing equations for the line
of sights near tangent to a caustic ring anywhere on the surface,
and estimated the lensing effects near its cusps. For a
cosmological axion caustic ring, the magnification may range
between $3\%$ and $2800\%$ at the outer cusp, and between $2\%$
and $46\%$ at the nonplanar cusps. For a cosmological WIMP caustic
ring, on the other hand, the magnification may range between $3\%$
and $28\%$ at the outer cusp, and between $2\%$ and $5\%$ at the
nonplanar cusps. As pointlike background sources pass behind the
caustics of foreground halos, the time scales for brightness to
change is about an hour for the axion folds and about a year for
the WIMP folds. Hence, depending on the strength of the observed
effect and the time scale for brightness change, it may even be
possible to discriminate between the axions and the WIMPs.

\section*{Acknowledgments}
This research was supported by European Union Commission Marie
Curie Fellowship FP-6-012679.

%\begin{thebibliography}{000} %for 3 digits
%\begin{thebibliography}{00}  %for 2 digits


\begin{thebibliography}{0}    %for 1 digit

\bibitem{Natural}
A thorough review can be found in J. F. Nye, {\it Natural Focusing
and Fine Structure of Light} (Institute of Physics Publishing,
Bristol, 1999).
\bibitem{Malin}
D. F. Malin and D. Carter, {\it Nature} {\bf 285}, 643 (1980).
\bibitem{Ripples}
L. Hernquist and P. J. Quinn, {\it Astrophys. J.} {\bf 312}, 1
(1985); E. Bertschinger, {\it Astrophys. J. Suppl.} {\bf 58}, 39
(1985); J. Binney and S. Tremaine, {\it Galactic Dynamics}
(Princeton University Press, 1987); A. Natarajan and P. Sikivie,
{\it Phys. Rev.} {\bf D72}, 083513 (2005).
\bibitem{Spergel}
D. N. Spergel et al., astro-ph/0603449.
\bibitem{Ipser}
P. Sikivie and J. R. Ipser, {\it Phys. Lett.} {\bf B291}, 288
(1992).
\bibitem{stw1}
P. Sikivie, I. Tkachev and Y. Wang, {\it Phys. Rev. Lett.} {\bf
75}, 2911 (1995).
\bibitem{zel}
Y. B. Zel'dovich, {\it Astrofizika} {\bf 6}, 319 (1970); {\it
Astron. Astrophys.} {\bf 5}, 84 (1970).
\bibitem{stw2}
P. Sikivie, I. Tkachev and Y. Wang, {\it Phys. Rev.} {\bf D56},
1863 (1997)
\bibitem{cr}
P. Sikivie, {\it Phys. Lett.}  {\bf B432}, 139 (1998).
\bibitem{sing}
P. Sikivie, {\it Phys. Rev.} {\bf D60}, 063501 (1999).
\bibitem{tre}
S. Tremaine, {\it Mon. Not. R. Astron. Soc.} {\bf 307}, 877
(1999).
\bibitem{Mahdavi}
A. Mahdavi, N. Trentham and R. B. Tully, astro-ph/0506737; R. B.
Tully, astro-ph/0509482.
\bibitem{KS}
W. Kinney and P. Sikivie, {\it Phys. Rev.} {\bf D61}, 087305
(2000).
\bibitem{IRAS}
P. Sikivie, {\it Phys. Lett.} {\bf B567}, 1 (2003).
\bibitem{Hogan}
C. Hogan,  {\it Astrophys. J.} {\bf 527}, 42 (1999).
\bibitem{lensing}
C. Charmousis, V. Onemli, Z. Qiu and P. Sikivie, {\it Phys. Rev.}
{\bf D67}, 103502 (2003).
%%CITATION = PHRVA,D67,103502;%%
\bibitem{Thesis}
Vakif K. Onemli, astro-ph/0401162, Ph. D. thesis, to be published
by the Nova Science Publishers, Inc., New York.
%%CITATION = ASTRO-PH 0401162;%%
\bibitem{probing}
R. Gavazzi, R. Mohayaee and B. Fort, {\it Astron. Astrophys.} {\bf
445}, 43 (2006); R. Mohayaee, S. Colombi, B. Fort, R. Gavazzi, S.
Shandarin and J. Touma, astro-ph/0510575.
\bibitem{Vakif}
V. K. Onemli, {\it Phys. Rev.} {\bf D74}, 123010 (2006),
astro-ph/0510414.
\bibitem{Wick}
P. Sikivie and S. Wick, {\it Phys. Rev.} {\bf D66}, 023504 (2002).
\bibitem{Chang}
S. Chang, C. Hagmann and P. Sikivie, {\it Phys. Rev.} {\bf D59},
023505 (1999).
\bibitem{FG}
J. A. Fillmore and P. Goldreich, {\it Astrophys. J.} {\bf 281}, 1
(1984); E. Bertschinger, {\it Astrophys. J. Suppl.} {\bf 58}, 39
(1985).
\bibitem{KK}
K. K. Nandi, Y.-Z. Zhang and A. V. Zakharov, {\it Phys. Rev.} {\bf
D74}, 024020 (2006); M.~Fairbairn, astro-ph/0511085.
\bibitem{Popowski}
C. Thomas et al., {\it Astrophys. J.} {\bf 631}, 906 (2005).

\end{thebibliography}
\end{document}